\documentstyle[aps,psfig,floats]{revtex}

\begin{document}
\draft 

\twocolumn[
\hsize\textwidth\columnwidth\hsize\csname@twocolumnfalse\endcsname

\title{Phenomenological interpretations of the ac Hall effect in the
normal state of ${\rm YBa_2Cu_3O_7}$}

\author{Anatoley T. Zheleznyak\cite{Zheleznyak}, Victor
M. Yakovenko\cite{Yakovenko}, and H. D. Drew\cite{Drew}}
\address{Department of Physics and Center for Superconductivity
Research, University of Maryland, College Park, Maryland 20742}
\author{I. I. Mazin}
\address{Computational Science and Informatics, George Mason
University and Naval Research Laboratory, 4555 Overlook Ave.,
Washington DC 20375}

\date{{\bf cond-mat/9706029}, posted June 3, 1997, revised December 8, 1997}

\maketitle

\begin{abstract}

ac and dc magnetotransport data in the normal state of ${\rm
YBa_2Cu_3O_7}$ are analyzed within Fermi-liquid and non-Fermi-liquid
models. In the Fermi-liquid analysis we use the Fermi surface deduced
from band-structure calculations and angular-resolved photoemission
experiments and assume that the electron relaxation rate varies over
the Fermi surface. The non-Fermi-liquid models are the two-dimensional
Luttinger liquid model and the charge-conjugation-symmetry model. We
find that the existing experimental data can be adequately fit by
any of these models. This work provides a framework for the analysis
of experiments that may discriminate among these models.

\end{abstract}

]

%%%%%%%%%%%%%%%%%%%%%%%%%%%%%%%%%%%%%%%%%%%%%%%%%%%%%%%%%%%%%%%%%%%%%%
\section{Introduction}
\label{sec:intro}
%%%%%%%%%%%%%%%%%%%%%%%%%%%%%%%%%%%%%%%%%%%%%%%%%%%%%%%%%%%%%%%%%%%%%%

\begin{figure}
\centerline{\psfig{file=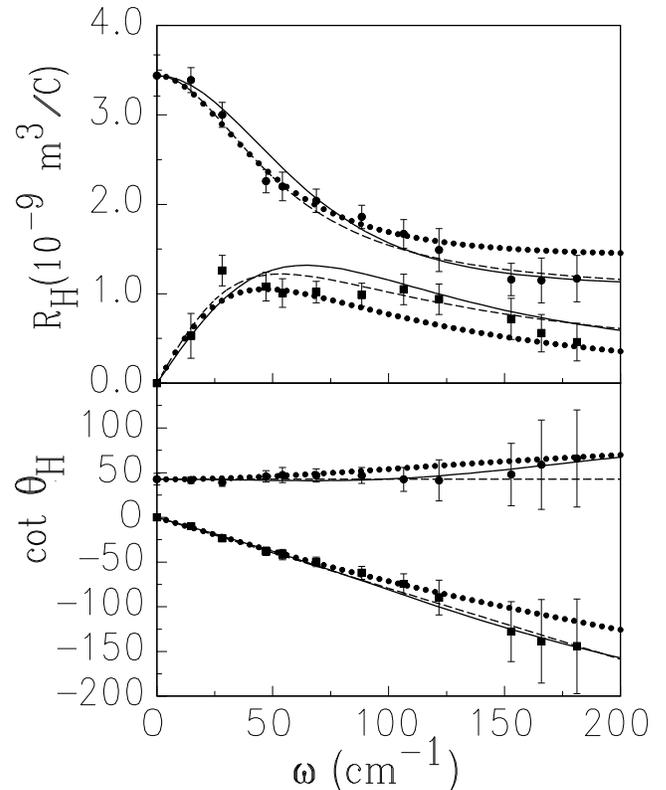,width=\linewidth}}
\caption{Frequency dependence of real and imaginary parts of the
inverse Hall angle $\cot \theta_H$ (lower panel) and the Hall
coefficient $R_H$ (upper panel).  Experimental values from Ref.\
{\protect \cite{Drew96}} are represented by the solid circles (real
parts) and the solid squares (imaginary parts).  The solid, dashed,
and dotted curves show the frequency dependences generated by the
additive two-$\tau$ model (Eqs.\ {\protect
(\ref{sxx2t})--(\ref{tau12tilde})}), by the multiplicative two-$\tau$
model (Eqs.\ {\protect (\ref{Anderson2tau}) and (\ref{taugeneral})}),
and by the charge-conjugation model (Eqs.\ {\protect
(\ref{sxxCol})--(\ref{gammaCol})}), respectively.}
\label{RHcotH}
\end{figure}

The simplest model of the electron transport in metals is the Drude
model, where all relaxation processes are described by a single
relaxation time $\tau$ \cite{Abrikosov}. This model fails to describe
the transport properties of high-$T_c$ superconductors in the normal
state. Evidence of the failure comes from comparison of the temperature
($T$) dependence of the in-plane resistivity $\rho_{xx}=1/\sigma_{xx}$
and the inverse Hall angle $\cot \theta_H = \sigma_{xx}/\sigma_{xy}$,
where $\sigma_{xx}$ and $\sigma_{xy}$ are the longitudinal and the Hall
components of the conductivity tensor. In the Drude model, both
$\rho_{xx}$ and $\cot \theta_H$ are proportional to the scattering rate
$\tau^{-1}$, and therefore should have the same temperature dependence.
Experiments, however, show a linear temperature dependence for
$\rho_{xx}$ and $T^2$ dependence for $\cot \theta_H$
\cite{Ong91,Ong92,Cooper92,Mihaly92,Cieplak92,Takagi92,Hwang94}.
Additional evidence comes from the frequency ($\omega$) dependence of
the Hall coefficient $R_H=\sigma_{xy}/(H\sigma_{xx}\sigma_{yy})$ and the
inverse Hall angle $\cot \theta_H$ in ${\rm Y Ba_2 Cu_3 O_7}$ thin films
\cite{Drew96}. The experimental data is shown in the upper and lower
panels of Fig.\ \ref{RHcotH}, where the solid circles represent the real
parts (Re) and the solid squares represent the imaginary parts (Im) of
$R_H(\omega)$ and $\cot \theta_H(\omega)$. While the inverse Hall angle
has a frequency dependence consistent with the Drude model: ${\rm
Re}\cot \theta_H(\omega)={\rm const}$ and ${\rm Im}\cot \theta_H(\omega)
\propto \omega$ (the lower panel in Fig.\ \ref{RHcotH}), the Hall
coefficient $R_H(\omega)$ exhibits non-Drude behavior (the upper panel
in Fig.\ \ref{RHcotH}). Indeed, in the single-relaxation-time model
${\rm Re} R_H$ is frequency independent, and ${\rm Im} R_H\equiv0$;
whereas experimentally ${\rm Re} R_H$ changes by a factor of $3$ from
$\omega=0$ to $\omega=200\;{\rm cm}^{-1}$, and ${\rm Im}
R_H(\omega)\not\equiv0$ \cite{Parks97}.

In order to explain different temperature dependences of $\rho_{xx}$
and $\cot \theta_H$, Anderson \cite{Anderson91} suggested that the
transport properties of cuprates are governed by two distinct
relaxation times $\tau_{\rm tr} \propto T^{-1}$ and $\tau_H \propto
T^{-2}$. The first relaxation time controls the longitudinal
conductivity, whereas the {\em product} of the two times determines
the Hall conductivity:
\begin{equation}
   \sigma_{xx} = \frac{\omega_p^2\tau_{\rm tr}}{4\pi}, \quad
   \sigma_{xy} = \frac{\omega_p^2\omega_H\tau_{\rm tr} \tau_H}{4\pi}.
\label{Anderson2tau}
\end{equation}
In a conventional model, the coefficients $\omega_p$ and $\omega_H$ in
Eq.\ (\ref{Anderson2tau}) would be the plasma and the cyclotron
frequencies, but here they are proposed to be related to the dynamical
response of the quasiparticles of the two-dimensional (2D) Luttinger
liquid.  Generalizing Eq.\ (\ref{Anderson2tau}) to a finite frequency
in a conventional manner:
\begin{equation}
\tau_j^{-1} \to \tau_j^{-1}-i\omega, \quad j={\rm tr},H,
\label{taugeneral}
\end{equation}

Kaplan {\it et al.} \cite{Drew96} produced a good quantitative fit of
their magneto-optical data \cite{Romero92a}. However, a microscopic
justification of Anderson's hypothesis is problematic. Anderson argued
that the two relaxation times originate from two different
quasiparticles, holons and spinons, in the Luttinger-liquid picture of
two-dimensional electron gas \cite{Anderson91}. However, distinct
kinetic equations for spinons and holons producing two different
relaxation times have not been demonstrated explicitly. In a heuristic
discussion, the two times were introduced as different coefficients in
front of the electric and the Lorentz forces in the stationary
Boltzmann equation for the {\em electrons} \cite{Harris95}. However,
this procedure violates Lorentz invariance between the electric and
Lorentz forces and contradicts the electron equation of motion, which
uniquely determines the force term in the Boltzmann equation. It is
problematic to write the kinetic equation of Ref.\ \cite{Harris95} as
a time-dependent Boltzmann equation, because it is not clear whether
$\tau_{\rm tr}$ or $\tau_H$ should be placed in front of the time
derivative of the electron distribution function \cite{note1}. Lee and
Lee \cite{PALee} studied the Hall effect in a holon-spinon liquid and
found a frequency-independent $R_H(\omega)$, the same result as in
Refs.\ \cite{Romero92a} and \cite{note1}, which does not agree with
experiment \cite{Drew96}. Abrahams \cite{Abrahams96} demonstrated that
Eq.\ (\ref{Anderson2tau}) can be derived from the Boltzmann equation
for the electrons if the scattering integral has two different
relaxation times for the electron velocities parallel and
perpendicular to the applied electric field. However, this assumption
cannot be valid in linear-response theory, where the scattering
integral does not depend explicitly on the infinitesimal external
electric field. Moreover, Kotliar, Sengupta, and Varma
\cite{Kotliar96} proved in general that it is not possible to obtain
the multiplicative rule (\ref{Anderson2tau}) from a Boltzmann equation
for the electrons. They observed that skew scattering, if diverging as
$1/T$, would produce the temperature dependence of the dc
magnetotransport. However, their model predicts that $R_H(\omega)$
does not depend on $\omega$, contradicting the ac experiment
\cite{Drew96}. Coleman, Schofield, and Tsvelik \cite{Coleman96}
proposed that the true quasiparticles may have even and odd
charge-conjugation symmetry, and their relaxation is characterized by
two different rates $\Gamma_f$ and $\Gamma_s$. That assumption
requires a Bogolyubov transformation of electrons into Majorana
fermions, which implies some sort of Cooper pairing with the total
momentum of twice the Fermi momentum. Physical justification for the
charge-conjugation symmetry is not clear. Lange \cite{Lange97}
demonstrated that the frequency dependence of the Hall constant
$R_H(\omega)$ corresponding to the ansatz (\ref{Anderson2tau}) and
(\ref{taugeneral}) can be naturally obtained within the
memory-function formalism. However, a microscopic calculation within
this formalism is not available yet.

An alternative phenomenological explanation for the temperature
dependences of $\rho_{xx}$ and $R_H$ within standard Fermi-liquid
theory with the electron scattering rate varying over the Fermi
surface was proposed in Refs.\ \cite{Cooper92,Mihaly92}. A
conceptually similar approach \cite{Allen92} was quantitatively
successful in describing the Hall coefficient of simple cubic metals.
In the approach of Refs.\ \cite{Cooper92,Mihaly92}, the Fermi surface
of ${\rm YBa_2Cu_3O_7}$ is assumed to have a specific geometry: large
flat regions and sharp corners. The electron scattering rate is
assumed to have linear temperature dependence on the flat parts of the
Fermi surface and quadratic in the corners. With the appropriate
choice of parameters, the longitudinal conductivity is dominated by
the contribution from the flat regions, whereas the sharply curved
corners control the Hall conductivity, and the model approximately
yields the required temperature dependences, $\sigma_{xx} \propto T$
and $\sigma_{xy} \propto T^3$ \cite{Cooper92,Mihaly92}. A number of
papers examined this approach microscopically: the linear temperature
dependence of $1/\tau$ within the nested-Fermi-liquid model
\cite{Ruvalds90} and the distribution of $1/\tau$ over the Fermi
surface within the model of antiferromagnetic fluctuations
\cite{Wheatley95,Hlubina95,Pines96,Dahm96,Kampf}. The degree of
quantitative agreement between the experiment and the model by
Stojkovi\'c and Pines was debated in Refs.\ \cite{Anderson97}.
Using a very different microscopic approach, the formalism of slave
bosons, S\'a de Melo, Wang, and Doniach \cite{Doniach92} studied the
distribution of $\tau$ over an effective Fermi surface of the $t$-$J$
model and calculated $R_H$ using an equation equivalent to our Eqs.\ 
(\ref{sxx}) and (\ref{sxy}).  An indirect experimental evidence for
the anisotropy of $\tau$ was found in the angular dependence of
magnetoresistance in ${\rm Tl_2Ba_2CuO_6}$ \cite{Hussey96}. A strong
variation of $\tau$ over the Fermi surface and the so-called ``hot
spots'' occur not only in the theoretical models of magnetotransport
in high-$T_c$ superconductors, but also in organic
quasi-one-dimensional conductors \cite{Zheleznyak95a}.

Recent measurements of thermopower \cite{Clayhold} produced an
evidence that the two lifetimes exist in the cuprate metals even
without magnetic field. That would eliminate the theoretical models
where the second lifetime has a purely magnetic origin, such as the
gauge model \cite{PALee} and the skew scattering model
\cite{Kotliar96}. The thermopower experiment poses a challenge for
Anderson's model \cite{Anderson91}, where the two lifetimes are
associated with the processes normal and tangential to the Fermi
surface, because thermopower involves only the normal processes.  On
the other hand, the charge-conjugation and the Fermi-liquid models,
where the two lifetimes exist irrespective of magnetic field, are
compatible with the experiment.

In this paper, we generalize the phenomenological Fermi-liquid
approach of Refs.\ \cite{Cooper92,Mihaly92} to finite frequencies. We
fit $R_H(\omega)$ and $\cot \theta_H(\omega)$ from Ref.\ \cite{Drew96}
using a model in which different parts of the Fermi surface are
characterized by two distinct relaxation times $\tau_1$ and
$\tau_2$. Since the two parts of the Fermi surface contribute
additively to the conductivity tensor, we call this model the additive
two-$\tau$ model. The additive law (\ref{sxx2t}) and (\ref{sxy2t}) of
this model is in the contrast to the multiplicative law
(\ref{Anderson2tau}) of Anderson's model. The additive law follows
naturally from the Boltzmann equation for the electrons, whereas the
multiplicative law has not been derived microscopically.

In our model, we assume that the relaxation times $\tau_1$ and
$\tau_2$ themselves do not depend on frequency. This assumption does
not contradict the experimental fact that $1/\tau \sim \omega$ at
high frequencies \cite{Thomas88,Schlesinger}, because we restrict our
fits to the relatively low-frequency range $\omega \le 200$ ${\rm
cm}^{-1}$, where the frequency dependence of $\tau$ is not yet
significant.

%%%%%%%%%%%%%%%%%%%%%%%%%%%%%%%%%%%%%%%%%%%%%%%%%%%%%%%%%%%%%%%%%%%%%%
\section{Fitting $\sigma_{\lowercase{xx}}(\omega)$ and
$\sigma_{\lowercase{xy}}(\omega)$ in an additive two-$\tau$ model}
\label{sec:2tau}
%%%%%%%%%%%%%%%%%%%%%%%%%%%%%%%%%%%%%%%%%%%%%%%%%%%%%%%%%%%%%%%%%%%%%%

We consider a layered electronic system consisting of two-dimensional
(2D) square lattices parallel to the $(x,y)$ plane and spaced along
the $z$ axis with the distance $d$. A weak magnetic field $H$ is
applied perpendicular to the planes. We neglect coupling between the
layers and assume that electrons form a 2D Fermi surface. Different
points on the Fermi surface can be labeled by $k_t$, the traverse
component of the electron wave vector. In general, the electron
relaxation time $\tau(k_t)$ and the Fermi velocity
$v(k_t)=\sqrt{v_x^2(k_t) +v_y^2(k_t)}$ may vary along the Fermi
surface. Within the conventional relaxation-time approximation for the
Boltzmann equation \cite{Abrikosov}, the components of the
frequency-dependent conductivity tensor are given by the following
equations:
\begin{eqnarray}
\sigma_{xx}(\omega) & = &  \frac{e^2}{2\pi^2 \hbar d}
     \oint \frac{v_x^2(k_t) \tilde {\tau}(k_t, \omega)}
     {v(k_t)}\, dk_t   \nonumber \\
& = & \frac{e^2}{(2\pi)^2 \hbar d}
     \oint v(k_t) \tilde {\tau}(k_t, \omega)\, dk_t,
\label{sxx} \\
\sigma_{xy}(\omega) & = &  \frac{e^3 H}{(2\pi\hbar)^2 c d} \oint dk_t
     \nonumber \\
& & {\bf e}_z \cdot \left[ {\bf v}(k_t) \tilde{\tau}(k_t,\omega)\times\;
     \frac{d \left[{\bf v}(k_t) \tilde{\tau}(k_t,\omega)\right]}{d k_t}
     \right],
\label{sxy} \\
\frac{1}{\tilde{\tau}(k_t,\omega)} & = & \frac{1}{\tau(k_t)} - i\omega,
\label{tauktilde}
\end{eqnarray}
where $\hbar$ is the Planck constant, $c$ is the speed of light, $e$
is the electron charge, ${\bf e}_z$ is a unit vector along the $z$
axis, and the integrals are taken over the Fermi surface.  Equation
(\ref{sxy}) is a finite-frequency generalization of Ong's formula
\cite{Ong91a}, which expresses the Hall conductivity of a 2D system in
terms of the area enclosed by the mean-free-path curve. Equation
(\ref{sxy}) applies when magnetic field is weak: $\omega_c\tau\ll1$,
where $\omega_c=(2\pi eH/\hbar c)[\oint dk_t/v(k_t)]^{-1}$ is the
cyclotron frequency corresponding to the electron motion around the
Fermi surface \cite{Abrikosov}.

Although in general the scattering rate should be continuously
distributed over the Fermi surface, to simplify analysis we consider a
model in which the Fermi surface is divided into two regions
characterized by different relaxation times $\tau_1$ and
$\tau_2$. With this assumption, Eqs.\ (\ref{sxx}) and (\ref{sxy}) can
be parametrized as follows:
\begin{eqnarray}
&&  \sigma_{xx}(\omega) = \frac{\omega_p^2}{4\pi}\;
    [a_1\tilde{\tau}_1(\omega) + a_2 \tilde{\tau}_2(\omega)],
    \quad a_1+a_2=1,
\label{sxx2t} \\
&&  \sigma_{xy}(\omega) = \frac{\omega_p^2 \omega_H}{4\pi}\;
    [b_1 \tilde{\tau}_1^2(\omega)+ b_2 \tilde{\tau}_2^2(\omega)],
    \quad b_1+b_2=1, \label{sxy2t} \\
&&  \frac{1}{\tilde{\tau}_{1,2}(\omega)}=\frac{1}{\tau_{1,2}}-i\omega,
\label{tau12tilde}
\end{eqnarray}
where
\begin{eqnarray}
&&   \omega_p^2 = \frac{e^2}{\pi \hbar d} \oint v(k_t) \,dk_t,
     \label{omp} \\
&&   \omega_H = \frac{eH}{\hbar c} \oint {\bf e}_z \cdot
      \big[{\bf v}(k_t) \times d{\bf v}(k_t) \big] \;
     \bigg / \oint v(k_t)\, dk_t, \label{omH} \\
&&    a_1 = \int_1 v(k_t) \; dk_t\; \bigg / \oint v(k_t) \, dk_t,
      \label{a1} \\
&&    b_1 =  \nonumber \\
&&    \int_1 {\bf e}_z \cdot \big[{\bf v}(k_t) 
      \times d{\bf v}(k_t)\big]\;
      \bigg / \oint {\bf e}_z \cdot \big[{\bf v}(k_t) 
      \times d{\bf v}(k_t) \big].
      \label{b1}
\end{eqnarray}
In Eqs.\ (\ref{a1}) and (\ref{b1}), $\int_1$ denotes integration over
the part of the Fermi surface with the relaxation time $\tau_1$. Equation
(\ref{omp}) defines the plasma frequency $\omega_p$ of the electron gas.
The frequency $\omega_H$ in Eq.\ (\ref{omH}) is proportional to the
magnetic field $H$, however is does not coincide with the cyclotron
frequency $\omega_c$ except for a circular Fermi surface. Equations
(\ref{sxx2t})--(\ref{tau12tilde}) represent the general form for
$\sigma_{xx}(\omega)$ and $\sigma_{xy}(\omega)$ in the additive
two-$\tau$ model and in this general case correspond to a conventional
two-band model. The equations contain six parameters: the prefactors
$\omega_p$ and $\omega_H$, the weights $a_1$ and $b_1$, and the
scattering times $\tau_1$ and $\tau_2$. The dimensional parameters
$\omega_p$ and $\omega_H$ determine the overall scale of $\sigma_{xx}$
and $\sigma_{xy}$, whereas the time $\tau_1$ sets the overall frequency
scale. The remaining three dimensionless parameters $a_1$, $b_1$, and
$\xi=\tau_2/\tau_1$ determine the shape of the frequency dependence. The
frequency $\omega$ appears in Eqs.\ (\ref{sxx2t}) and (\ref{sxy2t}) only
through the effective times $\tilde{\tau}_1(\omega)$ and
$\tilde{\tau}_2(\omega)$ defined in Eq.\ (\ref{tau12tilde}). At high
frequencies $\omega \gg 1/\tau_{1,2}$, the two effective times coincide:
$\tilde{\tau}_1(\omega) = \tilde{\tau}_2(\omega) = -i\omega$, and the
model reduces to a simple Drude model. (The discussion of the
high-frequency limit $\omega \gg 1/\tau_{1,2}$ here and in the next
paragraph is given only to illustrate the mathematical properties of the
functions involved. It should not be applied literally to the
high-frequency experiment, where the relaxation time $\tau_1$ itself
starts to depend on frequency: $1/\tau_1 \sim \omega$
\cite{Thomas88,Schlesinger}.)

The additive two-$\tau$ model can describe the observed frequency
dependence of the Hall coefficient $R_H = \sigma_{xy}/(H
\sigma_{xx}^2)$. As follows from Eqs.\
(\ref{sxx2t})--(\ref{tau12tilde}), ${\rm Re} R_H(\omega)$ starts at
one value at $\omega = 0$ and saturates to a different value at
$\omega = \infty$. While ${\rm Im} R_H = 0$ at $\omega = 0$ and
$\omega = \infty$, it is in general nonzero at other frequencies. This
is the qualitative behavior of the experimental data shown in the
upper panel of Fig.\ \ref{RHcotH}. The ratio of the low- and
high-frequency limits,
\begin{equation}
   \frac{{\rm Re} R_H(\omega=0)}{{\rm Re} R_H(\omega=\infty)} =
   \frac{b_1\tau_1^2+b_2 \tau_2^2}{(a_1\tau_1+a_2 \tau_2)^2},
   \label{ReRHr}
\end{equation}
can be fitted to the experimental value. On the other hand, in the
additive two-$\tau$ model, $\cot \theta_H(\omega)$ cannot have an
exactly linear dependence on $\omega$. However, it can have a
dependence that is close to linear for special choices of the model
parameters.  This is in contrast to the multiplicative model (Eqs.\
(\ref{Anderson2tau}) and (\ref{taugeneral})), which automatically
produces ${\rm Re}\cot\theta_H(\omega)={\rm const}$ and ${\rm
Im}\cot\theta_H(\omega) \propto \omega$. This important distinction
between the two models can be tested by measuring $\cot
\theta_H(\omega)$ at different temperatures and verifying whether a
linear frequency dependence exists at all temperatures.

We now discuss the analysis of the experimental data of Ref.\
\cite{Drew96} in terms of the additive two-$\tau$ model
(\ref{sxx2t})--(\ref{tau12tilde}). Instead of fitting $\cot \theta_H$
and $R_H$ deduced from the experimental data by Kramers-Kronig
analysis, we fit the raw data shown in Fig.\ \ref{TTH} by solid
circles. The lower panel in this figure shows the transmission
spectrum ${\cal T}(\omega)$ in zero magnetic field, and the upper
panel gives the ratio of the transmittances ${\cal T}^+(\omega)$ and
${\cal T}^-(\omega)$ of circularly polarized light for two opposite
orientations of the magnetic field: $H=9$ T and $H=-9$ T.  The
transmission coefficient of a thin film of thickness $D$ is related to
conductivities (\ref{sxx2t}) and (\ref{sxy2t}) by the standard formula
that takes into account multiple reflections in the substrate
\cite{Drew96}:
\begin{equation}
    {\cal T}^{\pm}(\omega) = \frac{4\; n\; {\cal T}_{{\rm corr}}}
    {|1+n+Z_0 D (\sigma_{xx} \pm i\sigma_{xy})|^2}, \label{Tpm}
\end{equation}
where
\begin{eqnarray}
&&   {\cal T}_{{\rm corr}} = \frac{1+R_s}{1-R_s R^{\pm}(\omega)},
     \label{Tcorr} \\
&&   R_s=\bigg(\frac{n-1}{n+1}\bigg)^2 , \\
&&   R^{\pm}(\omega) = \bigg|\frac{1-n+Z_0 D (\sigma_{xx} \pm i\sigma_{xy})}
      {1+n+Z_0 D (\sigma_{xx} \pm i\sigma_{xy})}\bigg|^2, \label{Rpm}
\end{eqnarray}
$Z_0=4\pi/c$ is the impedance of free space, and $n$ is the substrate
refraction index. The superscripts $\pm$ in Eqs.\ (\ref{Tpm}),
(\ref{Tcorr}), and (\ref{Rpm}) refer to the magnetic field parallel
and antiparallel to the $z$ axis.

Because the additive two-$\tau$ model has many adjustable parameters,
we divide our fitting procedure into two stages and each stage into
several steps. At the first stage we treat all six parameters of the
two-$\tau$ model as independent variables. At this stage we examine
how well the general additive two-$\tau$ model fits the experimental
data. We consecutively optimize over the fitting parameters, leaving
the optimization over $a_1$ and $b_1$ for the last step. At the second
stage (Sec.\ \ref{sec:mapping}) we take into account that $a_1$ and
$b_1$ are related to the electron dispersion law via Eqs.\ (\ref{a1})
and (\ref{b1}) and perform optimization for a model Fermi surface of
${\rm Y Ba_2 Cu_3 O_7}$.

First we fit the transmittance spectrum in zero magnetic field, ${\cal
T}(\omega)$, which involves the parameters $a_1$, $\tau_1$, $\tau_2$,
and $\omega_p$, but not $b_1$ and $\omega_H$. For the fixed values of
the parameters $a_1 \le 1$ and $\xi=\tau_2/\tau_1 > 1$, we find the
value of $\omega_p(a_1,\xi,\tau_1)$ from ${\cal T}(\omega=0)$ and the
value of $\tau_1(a_1,\xi)$ by minimizing the mean-square deviation
between the theoretical and experimental data for ${\cal T}(\omega)$:
\begin{equation}
  \chi_1 = \sqrt{ \frac{1}{N} \sum_{i=1}^N \bigg|
  \frac{{\cal T}^{({\rm exp})}(\omega_i)-{\cal T}(\omega_i)}
  {{\cal T}^{({\rm exp})}(\omega_i)} \bigg|^2 }. \label{msdT}
\end{equation}
In Eq.\ (\ref{msdT}) the sum is taken over the $N$ experimental values
of frequencies, the superscript (exp) denotes the experimental data,
and ${\cal T}(\omega_i)$ refers to the theoretical values from the
additive two-$\tau$ model. Then we scan the values of the parameters
$a_1$ and $\xi$, and retain for further consideration only those sets
$(a_1, \xi)$ where $\chi_1 \le 5\%$ and $\xi \le 6$. The latter
condition is imposed on the grounds that the scattering rate variation
$\xi=\tau_2/\tau_1$ should not be too large, because a very strong
variation of $\tau$ would be hard to justify within the Fermi-liquid
picture.

\begin{figure}
\centerline{\psfig{file=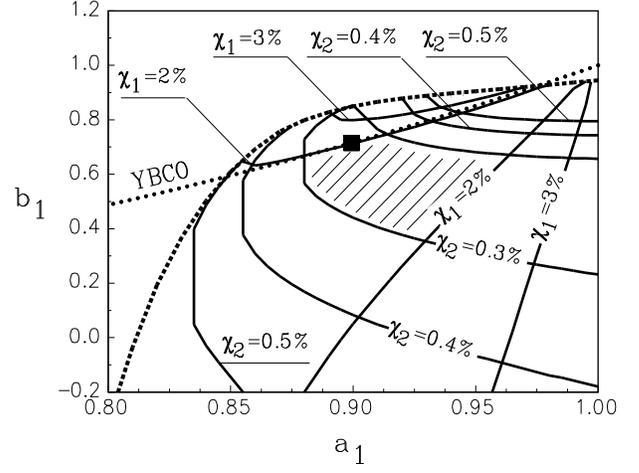,width=\linewidth,angle=-90}}
\caption{Contour plots of $\chi_1(a_1,b_1)$ {\protect (Eq.\
(\ref{msdT}))} and $\chi_2(a_1,b_1)$ {\protect (Eq.\ (\ref{msdTH}))}
shown by the solid lines.  Both $\chi_1(a_1, b_1)$ and $\chi_2(a_1,
b_1)$ are minimal in the shaded area: $\chi_1 < 2\%$ and $\chi_2<
0.3\%$.  Various constraints listed in Sec.\ {\protect \ref{sec:2tau}}
are satisfied below the dashed line. The dotted line, labeled ${\rm
YBCO}$, illustrates the relation between $a_1$ and $b_1$ for the ${\rm
CuO_2}$ bonding band of ${\rm Y Ba_2 Cu_3 O_7}$. The solid square
denotes the best mapping of the additive two-$\tau$ model onto the
${\rm CuO_2}$ bonding band of ${\rm Y Ba_2 Cu_3 O_7}$.}
\label{abcnpl}
\end{figure}

Next, for given values of $a_1$ and $\xi$, we determine the parameters
$\omega_H$ and $b_1$ in Eq.\ (\ref{sxy2t}) by matching the two
extremal points of the magnetic measurements: ${\rm Re} \cot \theta_H
(\omega = 0) = 43$ and ${\rm Re} R_H (\omega = 0)/ {\rm Re} R_H
(\omega = 200\;{\rm cm}^{-1})=3$.  For a fixed value of $a_1$,
$b_1(\xi)$ is found to first increase with increasing $\xi$ and then
decrease.  Thus, $b_1$ does not exceed a certain maximal value that
depends on $a_1$, and there exist two different values of $\xi$ that
generate the same value of $b_1$ below the maximum.  Again assuming
that the scattering rate variation should not be too large, we set
$\xi$ to the lower of these two values. With these restrictions, we
can map the pair of variables $(a_1, \xi)$ onto the pair $(a_1, b_1)$
and use the latter as the pair of independent variables.  To
characterize the quality of a fit in magnetic field, we calculate the
mean-square deviation between the experimental and theoretical data
for $r(\omega)={\cal T}^+(\omega)/{\cal T}^-(\omega)$:
\begin{equation}
   \chi_2=  \sqrt{ \frac{1}{N} \sum_{i=1}^N \bigg|
  \frac{r^{({\rm exp})}(\omega_i)-r(\omega_i)}
  {r^{({\rm exp})}(\omega_i)} \bigg|^2 }, \label{msdTH}
\end{equation}
and present the contour plots of $\chi_1(a_1,b_1)$ and
$\chi_2(a_1,b_1)$ in Fig.\ \ref{abcnpl}. The above listed constraints
are satisfied in the $(a_1,b_1)$ area below the dashed curve in Fig.\
\ref{abcnpl}.  The contour lines of $\chi_1(a_1, b_1)$ and
$\chi_2(a_1, b_1)$ intersect in a roughly orthogonal manner. The
optimal fit of the experimental curves for ${\cal T}(\omega)$ and
$r(\omega)$ is achieved in the shaded area in Fig.\ \ref{abcnpl},
where both deviations between the theoretical and experimental data
are minimal: $\chi_1(a_1, b_1) < 2 \%$ and $\chi_2(a_1, b_1) < 0.3
\%$. The latter value is smaller than the experimental error bars
approximately equal to $0.5 \%$.

It was noticed in Ref.\ \cite{Drew96} that the zero-field
transmittance ${\cal T}(\omega)$ can be well fitted with a simple
Drude model. According to Fig.\ \ref{abcnpl}, the simple Drude model
($a_1=1$) indeed provides a good fit for ${\cal T}(\omega)$ with
$\chi_1=3.5\%$, but the additive two-$\tau$ model provides a better
fit with $\chi_1(a_1, b_1) < 2 \%$. While this is natural because the
additive two-$\tau$ model has more fitting parameters, it does signify
that small corrections to the simple Drude model are required to
properly describe ${\cal T}(\omega)$.

%%%%%%%%%%%%%%%%%%%%%%%%%%%%%%%%%%%%%%%%%%%%%%%%%%%%%%%%%%%%%%%%%%%%%%
\section{Mapping $\tau_1$ and $\tau_2$ to the Fermi surface of
${\rm YB\lowercase{a}_2C\lowercase{u}_3O_7}$}
\label{sec:mapping}
%%%%%%%%%%%%%%%%%%%%%%%%%%%%%%%%%%%%%%%%%%%%%%%%%%%%%%%%%%%%%%%%%%%%%%

In this section we will incorporate the band structure of ${\rm
  YBa_2Cu_3O_7}$ into our fitting procedure for the additive
two-$\tau$ model. The band structure of ${\rm YBa_2Cu_3O_7}$ has been
calculated by several groups
\cite{Krakauer90,Mazin93,Mazin94,Andersen}. The Fermi surface of this
material is found to contain four sheets: the bonding and antibonding
bands originating from the two ${\rm CuO_2}$ planes, the ${\rm CuO}$
chain band, and a small column-shaped (the so-called ``stick'') pocket
originating from the ${\rm BaO}$ planes (see, for example, Fig.\ 2 in
Ref.\ \cite{Mazin94}). Photoemission measurements
\cite{Campuzano90,Olson92,Schabel97a,Schabel97b} agree qualitatively
with the theoretical results, although the stick pocket was reported
only in Ref.\ \cite{Campuzano90}. The chain band was observed in
positron annihilation experiments \cite{Peter}, and the stick pocket
has also been reported in the de Haas-van Alphen experiments
\cite{Haanappel}.  Unfortunately, it is very difficult to extract the
electron dispersion law in ${\rm YBa_2Cu_3O_7}$ from the photoemission
measurements quantitatively \cite{Schabel97b}. Nevertheless, a best
effort attempt was made by Schabel \cite{Schabel97c} for the
$\Gamma$--$S$ and $X$--$S$ directions in the ${\rm YBa_2Cu_3O_7}$
Brillouin zone. The band-structure calculations \cite{Mazin93,Mazin94}
give the following values of the Fermi wave vector and velocity along
the $\Gamma$--$S$ and $X$--$S$ directions for the bonding ${\rm
  CuO_2}$ band at $k_z=\pi/d$ \cite{dvsc}: $k_F^{(\Gamma-S)} = 0.9$
$\pi/a$, $k_F^{(X-S)} = 0.72$ $\pi/b$, $v^{(\Gamma-S)} = 0.6$ ${\rm
  eV} \cdot b/\hbar$, and $v^{(X-S)} = 1.2$ ${\rm eV} \cdot b/\hbar$,
where $a=3.8$ ${\rm \AA}$ and $b=3.9$ ${\rm \AA}$ are the lattice
constants of ${\rm YBa_2Cu_3O_7}$. These values agree with the data of
Schabel \cite{Schabel97c} within $10\%$ for $k_F$ and $25\%$ for
$v_F$.

\begin{figure}
\centerline{\psfig{file=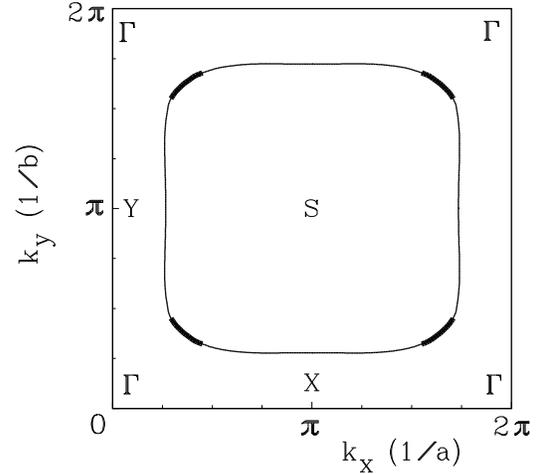,width=\linewidth,angle=-90}}
\caption{Fermi surface of the ${\rm CuO_2}$ bonding band of ${\rm
YBa_2Cu_3O_7}$ for $k_z=\pi/d$ according to the band-structure
calculations of Ref.\ {\protect \cite{Mazin94}}.  The regions denoted
by the thin lines have the short scattering time $\tau_1$, whereas the
bold regions have the long scattering time $\tau_2$.}
\label{FSkxky}
\end{figure}

This semiquantitative agreement with the photoemission experiment
encourages us to use the band calculations of Refs.\
\cite{Mazin93,Mazin94} in our analysis. We first calculate the
contributions of each band to $\sigma_{xx}$, $\sigma_{yy}$, and
$\sigma_{xy}$, assuming that all bands have a single relaxation time
$\tau$ and using Eqs.\ (\ref{sxx}) and (\ref{sxy}) integrated over
$k_z$ from $0$ to $2\pi/d$. The results are shown in Table
\ref{tab:bands}.  The values of total $\omega_{p,x}$, $\omega_{p,y}$,
and $R_H$ are in reasonable agreement with previous band-structure
calculations \cite{Allen88}. As follows from Table \ref{tab:bands},
the ${\rm CuO_2}$ bonding band gives maximal contribution to both the
longitudinal and Hall conductivities. Thus, to simplify the analysis,
we make the approximation of neglecting the contributions of the other
bands and consider only the bonding band. The Fermi surface of the
bonding band is shown in Fig.\ \ref{FSkxky} for $k_z=\pi/d$. It
contains the same geometrical features (large flat regions and sharp
corners) that were discussed for the additive two-$\tau$ model in
Refs.\ \cite{Cooper92,Mihaly92} (see Sec.\ \ref{sec:intro}). The
observed shape of the Fermi surface and the close values of the plasma
frequencies, $\omega_{p,x}^2= 4.1 \times 10^8$ ${\rm cm}^{-1}$ and
$\omega_{p,y}^2= 4.5 \times 10^8$ ${\rm cm}^{-1}$, confirm that the
bonding band has an approximate tetragonal symmetry. Strictly
speaking, the conductivities should be calculated for a given value of
$k_z$ using the 2D formalism of Sec.\ \ref{sec:2tau} and then
integrated over $k_z$. However, because the dispersion of the bonding
band in the $z$ direction is weak, we will simply use
$\varepsilon(k_x,k_y,k_z=\pi/d)$ in the equations of Sec.\
\ref{sec:2tau} as the 2D electron dispersion law. We select the value
$k_z=\pi/d$, since only for that value of $k_z$ the ${\rm CuO_2}$
bonding band does not hybridize with the ${\rm CuO}$ chain band
because of parity; thus, distortion of the plane band due to the chain
band is minimal.

In mapping the additive two-$\tau$ model considered in Sec.\
\ref{sec:2tau} onto the bonding band of ${\rm YBa_2Cu_3O_7}$, we must
address two basic questions. First, is it possible to divide the
Fermi surface of ${\rm YBa_2Cu_3O_7}$ in such a manner that the
dimensionless weights $a_1$ and $b_1$, calculated via Eqs.\ (\ref{a1})
and (\ref{b1}), have the values required by the additive two-$\tau$
model?  An answer to this question depends only on the variation of
the Fermi velocity ${\bf v}(k_t)$ over the Fermi surface, but not on
the overall scale of $v$. Second, do the values of the dimensional
parameters $\omega_p$ and $\omega_H$ for ${\rm YBa_2Cu_3O_7}$ agree
with those in the additive two-$\tau$ model?

We assign the shorter relaxation time $\tau_1$ to the large flat
regions, making them ``hot'' with respect to relaxation, and the
longer time $\tau_2$ to the corners (bold lines in Fig.\
\ref{FSkxky}), making them ``cold''. This assignment, which is
required to fit the ac and dc magnetotransport data for magnetic field
along the $c$ axis, is also consistent with the conclusions of the
transport experiments with the field in the $ab$ plane
\cite{Hussey96}. To find an optimal decomposition of the Fermi surface
into the hot and cold regions, we gradually increase the size of the
cold regions symmetrically with respect to the $\Gamma$--$S$ diagonal
and calculate the weights $a_1$ and $b_1$ of the contribution of the
hot regions to $\sigma_{xx}$ and $\sigma_{xy}$ from Eqs.\ (\ref{a1})
and (\ref{b1}). The weights $a_1$ and $b_1$ gradually decrease from
$1$ to $0$, which generates the dotted curve in Fig.\ \ref{abcnpl}
labeled ${\rm YBCO}$. The ${\rm YBCO}$ curve passes through the upper
part of the shaded area in Fig.\ \ref{abcnpl}, where the deviation of
the additive two-$\tau$ model from the experimental points is
minimal. The solid square in Fig.\ \ref{abcnpl} indicates the point of
the best mapping of the ${\rm YBa_2Cu_3O_7}$ Fermi surface onto the
additive two-$\tau$ model. The parameters of the additive two-$\tau$
model at this optimal point are given in Table
\ref{tab:param2tau}. The frequency dependences of ${\cal T}(\omega)$,
$r(\omega)$, $R_H(\omega)$, and $\cot \theta_H(\omega)$, generated in
the additive two-$\tau$ model with this set of parameters, are shown
by the solid lines in Figs.\ \ref{TTH} and \ref{RHcotH}. These lines
are in good agreement with the experimental points. Thus, the answer
to the first question formulated earlier in this section is positive:
The Fermi surface of ${\rm YBa_2Cu_3O_7}$ can be decomposed into the
hot and cold regions in such a manner that the shape of the frequency
dependences agrees well with the experiment.

Since the additive two-$\tau$ model with the parameters given in Table
\ref{tab:param2tau} fits the experimental data \cite{Drew96} very
well, we will refer to the values in Table \ref{tab:param2tau} as the
experimental values. The value of the plasma frequency,
$\omega_p^{({\rm exp})} = 10^4$ ${\rm cm}^{-1}$, in Table
\ref{tab:param2tau} is in reasonable agreement with the values
$\omega_{p,x}=10^4$ ${\rm cm}^{-1}$ and $\omega_{p,y}=1.6 \times 10^4$
${\rm cm}^{-1}$ found in previous measurements \cite{Basov}. (Unlike
\cite{Basov}, experiment \cite{Drew96} was performed on twinned
samples of ${\rm YBa_2Cu_3O_7}$, thus $\omega_p^{({\rm exp})}$ is an
average of $\omega_{p,x}$ and $\omega_{p,y}$. The difference between
$\omega_{p,x}$ and $\omega_{p,y}$ is mostly due to the ${\rm CuO}$
chains, which contribute predominantly to $\omega_{p,y}$, but not to
$\omega_{p,x}$.) The values $\omega_p^{({\rm exp})} = 10^4$ ${\rm
cm}^{-1}$ and $\omega_H^{({\rm exp})} = 1.7$ ${\rm cm}^{-1}$ from
Table \ref{tab:param2tau} differ considerably from the corresponding
values found in the band structure calculations for the ${\rm CuO_2}$
bonding band: $\omega_{p,x}^{({\rm bond})} = 2 \times 10^4$ ${\rm
cm}^{-1}$ and $\omega_H^{({\rm bond})} = 4$ ${\rm cm}^{-1}$ (see Table
\ref{tab:bands}). If $a_1$, $b_1$, $\tau_1$, and $\tau_2$ from Table
\ref{tab:param2tau} are assigned to the ${\rm CuO_2}$ bonding band
with $\omega_p$ and $\omega_H$ calculated in Table \ref{tab:bands},
the discrepancies between the calculated and measured values are as
follows: $\sigma_{xx}^{({\rm bond})}(\omega)/\sigma_{xx}^{({\rm
exp})}(\omega)= (\omega_{p,x}^{({\rm bond})}/\omega_p^{({\rm
exp})})^2=4$ and $R_H^{({\rm bond})}(\omega)/R_H^{({\rm
exp})}(\omega)= (\omega_H^{({\rm bond})}/\omega_H^{({\rm exp})})
(\omega_{p,x}^{({\rm bond})}/\omega_p^{({\rm exp})})^{-2}=0.6$. (The
ratios are frequency-independent, because the model matches the shape
of the experimental frequency dependence.) The discrepancy between the
calculated and experimental values of the plasma frequencies has been
noticed and discussed in literature \cite{Allen88,Basov}. The simplest
way to resolve this discrepancy is to assume that the Fermi velocity,
proportional to $\omega_p^2$ via Eq.\ (\ref{omp}), is uniformly
reduced by a factor of 4 due to many-body renormalization effects
coming from electron-phonon interaction or other correlation
effects. Indeed, a factor of two to four renormalization has been
deduced from a memory function analysis of infrared data by
Schlesinger {\it et al.}  \cite{Schlesinger}. However, the uniform
renormalization of the Fermi velocity cannot correct the discrepancy
in the Hall coefficient, since $R_H$ is not sensitive to the velocity
scaling factor (see Eqs.\ (\ref{sxx}) and (\ref{sxy})).  Thus, the
answer to the second question formulated earlier in this section is
negative: The overall scales of the transport coefficients calculated
for the band structure of ${\rm YBa_2Cu_3O_7}$ differ significantly
from the measured values.  Nevertheless, considering the crudeness of
our model assumptions (discontinuous distribution of $\tau$, neglected
contributions of other bands, and ignoring many-body renormalization
effects), the qualitative and semiquantitative agreement of the fits
indicates that the Fermi-liquid interpretation remains viable.

\begin{figure}
\centerline{\psfig{file=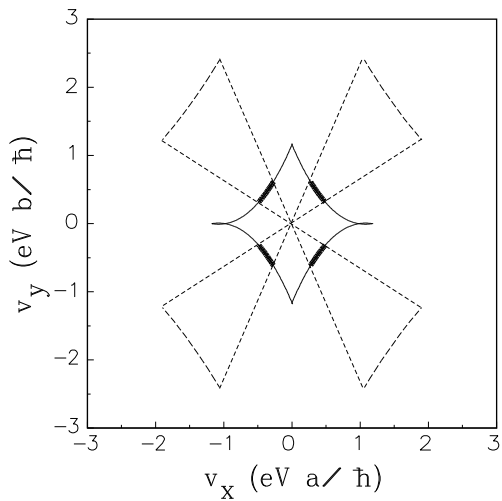,width=\linewidth,angle=-90}}
\caption{Solid lines: distribution of the Fermi velocity vector
($v_x,v_y)$ for the ${\rm CuO_2}$ bonding band of ${\rm YBa_2Cu_3O_7}$
according to the band-structure calculations of Ref.\ {\protect
\cite{Mazin94}} for $k_z=\pi/d$.  Bold lines: regions with the long
scattering time $\tau_2$.  Dashed lines: the Fermi velocity scaled by
the factor of $\xi=\tau_2/\tau_1=3.9$. The area enclosed by the dashed
and thin lines determines the zero-frequency Hall conductivity via
Ong's formula {\protect \cite{Ong91a}}.}
\label{FSvxvy}
\end{figure}

Every point $k_t$ on a 2D Fermi surface has a certain Fermi velocity
vector ${\bf v}(k_t)=(v_x(k_t),v_y(k_t))$. As the point $k_t$ moves
along the Fermi surface, the 2D vector ${\bf v}(k_t)$ traces a certain
curve in the 2D velocity space $(v_x,v_y)$. This curve is shown in
Fig.\ \ref{FSvxvy} by the solid curve for the bonding band of ${\rm
YBa_2Cu_3O_7}$. The cold regions of the Fermi surface are indicated by
the bold lines in Fig.\ \ref{FSvxvy}. Correspondingly, the
mean-free-path vector ${\bf l}(k_t)=\tau(k_t){\bf v}(k_t)$ also traces
a certain curve in the mean-free-path space $(l_x,l_y)$ (the {\bf
l} curve). It was shown by Ong \cite{Ong91a} that the Hall
conductivity $\sigma_{xy}$ of a 2D electron gas is proportional to the
area enclosed by the ${\bf l}$ curve (see Eq.\ (\ref{sxy})). To
illustrate the shape of the ${\bf l}$ curve, we multiply the Fermi
velocity in the cold regions by the factor $\xi=\tau_2/\tau_1=3.9$ and
present the scaled curve by the dashed lines in Fig.\
\ref{FSvxvy}. The ${\bf l}$ curve for the ${\rm YBa_2Cu_3O_7}$ bonding
band is the same (up to the overall factor $\tau_1$) as the curve
consisting of the dashed and thin lines in Fig.\ \ref{FSvxvy}. This
curve is qualitatively similar to the ${\bf l}$ curve found in Ref.\
\cite{Wheatley95} in the model of antiferromagnetic spin
fluctuations. On the other hand, our ${\bf l}$ curve differs from the
${\bf l}$ curve used in Refs.\ \cite{Cooper92,Mihaly92}, where it was
assumed the variation of ${\bf l}(k_t)=\tau(k_t){\bf v}(k_t)$ is
dominated by the variation of ${\bf v}(k_t)$. Contrarily, we find that
the variation of the scattering time $\tau(k_t)$ outweighs the
variation of the Fermi-surface velocity ${\bf v}(k_t)$. Since the area
of the sector enclosed by dashed lines in Fig.\ \ref{FSvxvy} is larger
than the area enclosed by the thin curves, the Hall conductivity at
zero frequency is dominated by the cold regions with the long
relaxation time $\tau_2$. In the opposite limit of high frequency, the
effective scattering times (\ref{tau12tilde}) become equal, and the
contributions of the hot and the cold regions of the Fermi surface to
the Hall conductivity $\sigma_{xy}$ become comparable.  In contrast,
the longitudinal conductivity $\sigma_{xx}$, which is proportional to
the mean-free-path average over the Fermi surface (\ref{sxx}), is
dominated by the hot regions of the Fermi surface in both high- and
low-frequency limits.

\begin{figure}
\centerline{\psfig{file=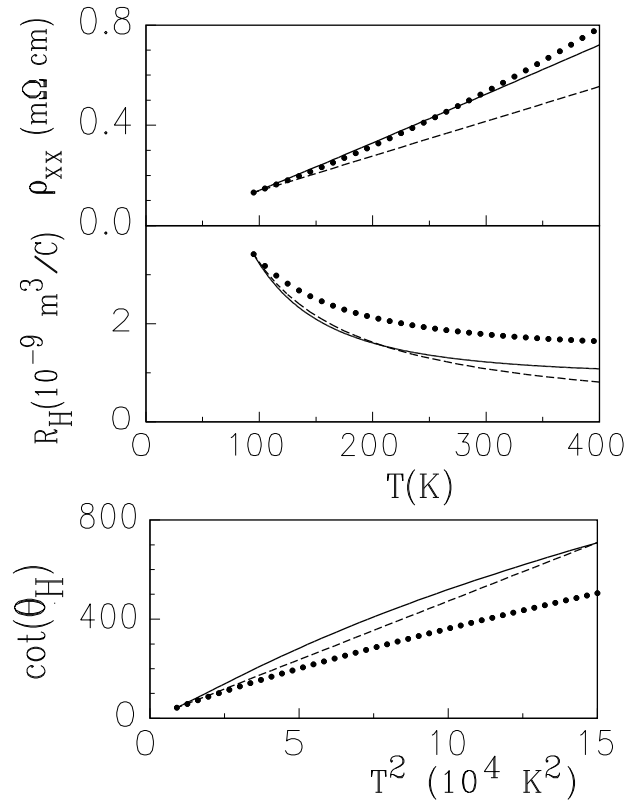,width=\linewidth}}
\caption{Temperature dependences of resistivity $\rho_{xx}$ (top
panel), the Hall coefficient $R_H$ (middle panel), and the inverse
Hall angle $\cot \theta_H$ at $H=9$ T (bottom panel), generated in the
additive and multiplicative two-$\tau$ models by assigning
$\tau_1^{-1}$, $\tau_{\rm tr}^{-1}$, $\Gamma_f \propto T$ and
$\tau_2^{-1}$, $\tau_H^{-1}$, $\Gamma_s \propto T^2$. The solid,
dashed, and dotted curves correspond to the additive, multiplicative,
and charge-conjugation models, respectively.}
\label{rxxcotHT}
\end{figure}

Following Ref.\ \cite{Mihaly92}, we assign linear and quadratic
temperature dependences to the scattering rates of the additive
two-$\tau$ model:
$$
\tau_1^{-1}=\eta T, \qquad \tau_2^{-1}=T^2/W.
$$
Using the values of $\tau_1^{-1}$ and $\tau_2^{-1}$ (see Table
\ref{tab:param2tau}) obtained by the fit of the ac Hall data to the
additive two-$\tau$ model and taking into account that the ac
measurements of Ref.\ \cite{Drew96} were taken at $T=95$ K, we find
$\eta=4.5$ and $W=82.5$ K.  Using these and the other parameters
listed in Table \ref{tab:param2tau}, we calculate the temperature
dependences $\rho_{xx}(T)$, $R_H(T)$, and $\cot \theta_H(T)$ for the
additive two-$\tau$ model and show them by the solid curves in Fig.\
\ref{rxxcotHT}. The top panel in Fig.\ \ref{rxxcotHT} demonstrates a
nearly linear temperature dependence of $\rho_{xx}(T)$. The bottom
panel shows the temperature dependence of $\cot \theta_H$. While not
exactly quadratic, it does resemble the experimental data of Refs.\
\cite{Ong92,Cooper92}. With the assumed temperature dependences for
the scattering times, we find that the two relaxation times become
approximately equal at $T=371$ K: $\tau_1(T=371\;{\rm K}) \approx
\tau_2(T=371$ ${\rm K})$. Thus, we may expect Drude-like behavior for
all frequency dependences at $T=371$ K, i.\ e., ${\rm Re}R_H(\omega)$
to be approximately frequency-independent and ${\rm Im}R_H(\omega)
\approx 0$.

In our discussion of the Hall effect, we have assumed the
low-magnetic-field limit $\omega_H\tau \ll 1$, where $\sigma_{xy}$ is
given by Eq.\ (\ref{sxy}) in terms of the distribution of the
scattering time and the Fermi velocity over the Fermi surface. Using
the parameters given in Table \ref{tab:param2tau}, we find that for
the moderate magnetic field of $9$ T used in experiment \cite{Drew96}
$\omega_H \tau_1 = 5.7 \times 10^{-3}$ and $\omega_H \tau_2 = 2.2
\times 10^{-2}$, thus the low-field condition $\omega_H\tau \ll 1$ is
indeed satisfied. On the other hand, in the strong-magnetic-field
limit $\omega_H\tau \gg 1$, the Hall coefficient is given by a
different formula \cite{Abrikosov} in terms of the concentration of
carriers (holes):
\begin{equation}
   R_H=\frac{V}{2ecS},
\label{RHsmf}
\end{equation}
where $S$ is the dimensionless fraction of the Brillouin zone enclosed
by the Fermi surface, and $V=abd$ is the unit-cell volume of the crystal
\cite{dvsc}. For the bonding band of ${\rm YBa_2Cu_3O_7}$, we find that
$S=0.51$, and the Hall coefficient in a strong magnetic field,
$R_H(\omega_H\tau \gg 1) = 1.1 \times 10^9$ ${\rm m^3/C} \approx 0.3 \,
R_H(\omega_H\tau \ll 1)$, is three times lower than in a weak magnetic
field. The three-times reduction of the dc $R_H$ from low to high
magnetic fields is approximately the same as the reduction of $R_H$ from
low to high frequencies (see Fig.\ \ref{RHcotH}). Both effects have the
common origin: The strong variation of $\tau$ over the Fermi surface,
essential for the Hall effect at low $\omega$ and low $H$, becomes
irrelevant for high $\omega$ or high $H$. Recent experiment in overdoped
$\rm Tl_2Ba_2CuO_{6+\delta}$ with $T_c\sim30$ K \cite{Boebinger} found a
decrease of $R_H$ in a very strong field of 60 T where
$\omega_H\tau=0.9$, in qualitative agreement with the theoretical
picture outlined above.

We can also consider mapping of the additive two-$\tau$ model onto two
distinct bands characterized by different relaxation times, for
example the ${\rm CuO_2}$ bonding band and the ${\rm BaO}$ stick
pocket.  Assuming for simplicity that the bands have parabolic
dispersion and using the parameters listed in Table
\ref{tab:param2tau}, we estimate the Fermi wave vectors, masses, and
plasma frequencies of the corresponding bands: $k_{F,1} = 0.84$
$\pi/a$ and $k_{F,2} = 0.14$ $\pi/a$, $m_1 = 6\, m$ and $m_2 =
1.7\,m$, $\omega_{p,1}^2 = 9 \times 10^7$ ${\rm cm}^{-1}$ and
$\omega_{p,2}^2 = 10^7$ ${\rm cm}^{-1}$, where $m$ is the
free-electron mass. The values of the Fermi momenta roughly agree with
the Fermi momenta of the ${\rm CuO_2}$ bonding band (see the beginning
of this Section) and the ${\rm BaO}$ stick pocket ($k_F^{({\rm
stick})}=0.11$ $\pi/a$ along the $S$--$X$ direction at
$k_z=\pi/d$) found in band-structure calculations \cite{Mazin94}, as
well as the values $0.12$ $\pi/a$ and $0.17$ $\pi/a$ found in de
Haas-van Alphen experiments \cite{Haanappel} for $k_F^{({\rm
    stick})}$. The value of $\omega_{p,1}$ differs significantly from
the plasma frequency of the bonding band by about the same factor that
we discussed earlier in this Section: $\omega_{p,1}^2 = 0.2\,
(\omega_{p,x}^{({\rm bond})})^2$, whereas $\omega_{p,2}$ is comparable
to the plasma frequency of the stick pocket, $\omega_{p,2}^2 = 1.25\,
(\omega_{p,x}^{({\rm stick})})^2$. However, de Haas-van Alphen
experiments \cite{Haanappel} give the value $7\,m$ for the mass of the
stick pocket, which strongly disagrees with the value required by the
two-$\tau$ model. The contribution of the stick pocket to the Hall
conductivity, 6\% according to the band-structure calculations (see
Table \ref{tab:bands}), is too small compared with the value
$b_2=1-b_1=29\%$ required by the two-$\tau$ model (see Table
\ref{tab:param2tau}). In other words, to fit experiment \cite{Drew96},
one needs a large and heavy Fermi surface, combined with a small and
light one. On the contrary, band-structure calculations
\cite{Mazin94}, as well as de Haas-van Alphen experiments
\cite{Haanappel}, evince the large and light bonding Fermi surface,
combined with the small and heavy stick Fermi surface. As a result,
mapping of the additive two-$\tau$ model onto the ${\rm CuO_2}$
bonding band and the ${\rm BaO}$ stick pocket does not appear to be
consistent with the ac and dc magnetotransport data.

%%%%%%%%%%%%%%%%%%%%%%%%%%%%%%%%%%%%%%%%%%%%%%%%%%%%%%%%%%%%%%%%%%%%%%%%%
\section{Fitting $\sigma_{\lowercase{xx}}(\omega)$ and
$\sigma_{\lowercase{xy}}(\omega)$ in multiplicative two-$\tau$ models}
\label{sec:Colopt}
%%%%%%%%%%%%%%%%%%%%%%%%%%%%%%%%%%%%%%%%%%%%%%%%%%%%%%%%%%%%%%%%%%%%%%%%%%

As mentioned in Sec.\ \ref{sec:intro}, the experimental data of Ref.\
\cite{Drew96} can be well fitted by the multiplicative two-$\tau$
model defined by Eqs.\ (\ref{Anderson2tau}) and (\ref{taugeneral}).
The charge-conjugation model \cite{Coleman96} has the same
multiplicative law for the Hall conductivity, but a different
expression for the longitudinal conductivity:
\begin{eqnarray}
&&  \sigma_{xx}(\omega) = \frac{\omega_p^2}
   {2\pi[\tilde {\Gamma}_f(\omega) + \tilde{\Gamma}_s(\omega)]},
\label{sxxCol} \\
&&  \sigma_{xy}(\omega)= \frac{\omega_p^2 \omega_H}
   {4\pi\tilde {\Gamma}_f(\omega) \tilde{\Gamma}_s(\omega)},
\label{sxyCol} \\
&&   \tilde {\Gamma}_j(\omega) = \Gamma_j-i\omega, \quad j=f,s.
\label{gammaCol}
\end{eqnarray}

\begin{figure}
\centerline{\psfig{file=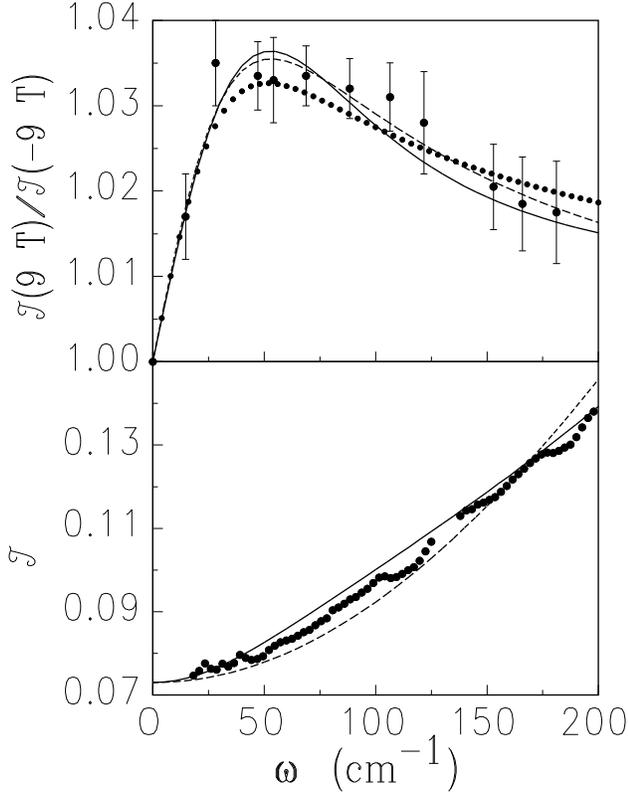,width=\linewidth}}
\caption{Far-infrared transmission spectra of a ${\rm Y Ba_2 Cu_3
O_7}$ thin film. The lower panel gives the frequency dependence of the
transmittance ${\cal T}(\omega)$ in zero magnetic field.  The upper
panel shows the ratio of the transmittances at $H=9$ and $H=-9$ T. The
solid circles represent the experimental values from Ref.\ {\protect
\cite{Drew96}}.  The solid curves show the frequency dependences
generated by the additive two-$\tau$ model {\protect
(\ref{sxx2t})--(\ref{tau12tilde})}.  The dashed and the dotted curves
in the upper panel represents the multiplicative two-$\tau$ model
(Eqs.\ {\protect (\ref{Anderson2tau}) and (\ref{taugeneral})}) and the
charge-conjugation model (Eqs.\ {\protect
(\ref{sxxCol})--(\ref{gammaCol})}), respectively.  The dashed curve in
the lower panel is obtained in the single-relaxation-time Drude
model.}
\label{TTH}
\end{figure}

In this section, we fit the experimental frequency dependences ${\cal
T}(\omega)$, $r(\omega)$, $R_H(\omega)$, and $\cot \theta_H(\omega)$
using both multiplicative two-$\tau$ models. Both models have four
phenomenological parameters: the prefactors $\omega_p$ and $\omega_H$
and the relaxation rates $\tau_{\rm tr}^{-1}$ and $\tau_H^{-1}$ or
$\Gamma_f$ and $\Gamma_s$. In both formalisms, fitting the zero-field
transmittance spectrum ${\cal T}(\omega)$ is equivalent to using a model
with a single relaxation time, which we have already studied in Sec.\
\ref{sec:2tau} as the special case $a_1=1$. That gives the following
values of the parameters: $\omega_p=9.2 \times 10^3$ ${\rm cm}^{-1}$,
$\tau_{\rm tr}^{-1}=(\Gamma_f+\Gamma_s)/2 = 185$ ${\rm cm}^{-1}$, and
$\chi_1=3.5\%$. The zero-field transmittance spectrum ${\cal
T}(\omega)$, generated by these parameters, is shown by the dashed curve
in the lower panel of Fig.\ \ref{TTH}. Then we find $\omega_H$ by
fitting $\cot\theta_H(\omega=0)=43$ and find the ratio of the relaxation
rates by minimizing the deviation $\chi_2$ (see Eq.\ (\ref{msdTH})). For
the model (\ref{Anderson2tau})--(\ref{taugeneral}), we find
$\tau_H^{-1}=54$ ${\rm cm}^{-1}$, $\omega_H^{-1}=1.3$ ${\rm cm}^{-1}$,
and $\chi_2=0.22\%$. For the charge-conjugation model
(\ref{sxxCol})--(\ref{gammaCol}), the values of the parameters are:
$\Gamma_f=322$ ${\rm cm}^{-1}$, $\Gamma_s=49$ ${\rm cm}^{-1}$,
$\omega_H=2$ ${\rm cm}^{-1}$, and $\chi_2=0.29\%$. The frequency
dependences of $R_H(\omega)$, $\cot\theta_H(\omega)$, and $r(\omega)$
generated with these sets of parameters are shown in Figs.\ \ref{RHcotH}
and \ref{TTH} by the dashed curves for the model
(\ref{Anderson2tau})--(\ref{taugeneral}) and by the dotted curves for
the model (\ref{sxxCol})--(\ref{gammaCol}). While both models are in
reasonably good agreement with the experimental data, the fit to
Anderson's model appears systematically better than the
charge-conjugation model. Assigning the following temperature
dependences to the scattering rates: $\tau_{\rm tr}^{-1}, \Gamma_f
\propto T$ and $\tau_H^{-1}, \Gamma_s \propto T^2$, we obtain the
temperature dependences $\rho_{xx}(T)$, $R_H(T)$, and $\cot\theta_H(T)$
shown in Fig.\ \ref{rxxcotHT} by the dashed lines for Anderson's model
and by the dotted lines for the charge-conjugation model. The
temperature dependences in all three models are relatively close to each
other and agree qualitatively with the experiment.

%%%%%%%%%%%%%%%%%%%%%%%%%%%%%%%%%%%%%%%%%%%%%%%%%%%%%%%%%%%%%%%%%%%%%%
\section{Conclusions}
\label{sec:concl}
%%%%%%%%%%%%%%%%%%%%%%%%%%%%%%%%%%%%%%%%%%%%%%%%%%%%%%%%%%%%%%%%%%%%%%

In this paper we have examined various phenomenological
interpretations of the ac Hall effect in the normal state of ${\rm
YBa_2Cu_3O_7}$. We have demonstrated that it is possible to fit the
magnetotransport data obtained in Ref.\ \cite{Drew96} within a
Fermi-liquid model, if different regions on the Fermi surface are
characterized by two different relaxation times (the additive
two-$\tau$ model). Mapping the additive two-$\tau$ model onto the
${\rm CuO_2}$ bonding band of ${\rm YBa_2Cu_3O_7}$ calculated in
Refs.\ \cite{Mazin93,Mazin94}, we find that the large flat regions of
the Fermi surface have a short relaxation time (are ``hot''), whereas
the sharp corners have a long relaxation time (are ``cold''). This
distribution of the relaxation times over the Fermi surface of the
${\rm CuO_2}$ bonding band allows us to fit the shape of the
experimental frequency dependences very well. On the other hand, there
are considerable discrepancies between the band-structure calculations
and the experiment in the overall magnitude of transport coefficients,
which can be partially resolved by including many-body renormalization
of the Fermi velocity.

We also find that the data of Ref.\ \cite{Drew96} can be well fitted
by the two unconventional, multiplicative two-$\tau$ models:
Anderson's 2D Luttinger-liquid model \cite{Anderson91} generalized to
finite frequencies by Kaplan {\it et al.} \cite{Drew96} and somewhat
less well by the charge-conjugation-symmetry model by Coleman,
Schofield, and Tsvelik \cite{Coleman96}. We conclude that the existing
experimental data does not permit a definitive discrimination between
these three models. Measurements of the frequency dependence of the ac
Hall effect at different temperatures would be very useful.  Since the
relaxation rates have different temperature dependences, the frequency
dependences of magnetotransport coefficients in the three models
should change with temperature differently, so that discrimination
between the different models may become possible.

%%%%%%%%%%%%%%%%%%%%%%%%%%%%%%%%%%%%%%%%%%%%%%%%%%%%%%%%%%%%%%%%%%%%%%%%%%%
\section*{Acknowledgments}
%%%%%%%%%%%%%%%%%%%%%%%%%%%%%%%%%%%%%%%%%%%%%%%%%%%%%%%%%%%%%%%%%%%%%%%%%%
We are grateful to M.~C.~Schabel for sending us preprint \cite{Schabel97b} 
and unpublished data on the electron dispersion in ${\rm YBa_2Cu_3O_7}$ 
and to J.~M.~Harris and G.~S.~Boebinger for useful discussions.
This work was supported by the David and Lucile Packard Foundation and by 
the NSF under Grants No. DMR-9417451 and DMR-9223217.
The work at NRL was supported by the Office of Naval Research.

%%%%%%%%%%%%%%%%%%%%%%%%%%%%%%%%%%%%%%%%%%%%%%%%%%%%%%%%%%%%%%%%%%%%%%%%%%%

\onecolumn

\begin{table}
\caption{\label{tab:bands} Contributions of different ${\rm
YBa_2Cu_3O_7}$ bands to the longitudinal and Hall conductivities in a
single-relaxation-time-$\tau$ model. As it is conventional in optics,
the frequencies are given in ${\rm cm}^{-1}$. The plasma frequencies
$\omega_{p,x}$ and $\omega_{p,y}$ are taken from Ref.\ {\protect
\cite{Mazin93}}.}
\begin{tabular}{|l||c|c|c|c|c|}
 & & Bonding & Antibonding & Chains & Stick \\
 & \raisebox{1.5ex}[0pt]{Total}
 & ${\rm CuO_2}$ band
 & ${\rm CuO_2}$ band
 & ${\rm CuO}$ band
 & ${\rm BaO}$ pocket
 \\ \hline \hline
$4\pi\sigma_{xx}/\tau=\omega_{p,x}^2$ & $7 \times 10^8$ ${\rm cm}^{-2}$ &
$59\%$ & $36\%$ & $4\%$ & $1\%$ \\ \hline
$4\pi\sigma_{yy}/\tau=\omega_{p,y}^2$ & $12\times 10^8$ ${\rm cm}^{-2}$ &
$37\%$ & $32\%$ & $30\%$ & $1\%$ \\ \hline
$4\pi\sigma_{xy}/\tau^2$ at $H=9$ T & $2 \times 10^9$ ${\rm cm}^{-3}$ &
$82\%$ & $40\%$ & $-28\%$ & $6\%$ \\ \hline
$R_H=\sigma_{xy}/(H\sigma_{xx}\sigma_{yy})$ & 
$0.16 \times 10^{-9}$ ${\rm m^3/C}$ & & & & \\ 
\end{tabular}
\end{table}
%\nopagebreak

\begin{table}
\caption{\label{tab:param2tau} Parameters of the additive two-$\tau$
model for the best mapping onto the ${\rm YBa_2Cu_3O_7}$ Fermi
surface.}
\begin{tabular}{|c|c|c|c|c|c|c|c|c|}
$a_1$ & $b_1$ & $\xi=\tau_2/\tau_1$ & $\tau_1^{-1}$ & $\tau_2^{-1}$ &
$\omega_p$ & $\omega_H$ at $H=9$ T & $\chi_1$ & $\chi_2$        \\ \hline
$0.9$ & $0.71$ & $3.9$ & $297$ ${\rm cm}^{-1}$ & $76$ ${\rm cm}^{-1}$ &
$10^4$ ${\rm cm}^{-1}$ & $1.7$ ${\rm cm}^{-1}$ & $2\%$  & $0.3\%$ \\
\end{tabular}
\end{table}


\begin{references}
\bibitem{dvsc} In Sec.\ {\protect \ref{sec:mapping}} we use the letter
$d$ to denote the lattice spacing of ${\rm YBa_2Cu_3O_7}$ in the $c$
direction, not the distance between ${\rm CuO_2}$ layers. This spacing
is usually denoted by the letter $c$, but we reserve that letter for
the speed of light.
\end{references}
\end{document}